\newcommand*{\ARXIV}{}%
\ifdefined\ARXIV
\else
\fi

\ifdefined\ARXIV
\documentclass{article}
\usepackage{arxiv}

\else
\documentclass[sigconf, screen, nonacm]{acmart}
\fi

\ifdefined\ARXIV
\else
\setcopyright{acmcopyright}
\copyrightyear{2021}
\acmYear{2021}
\acmDOI{10.1145/1122445.1122456}

\acmConference[ACM'21]{ACM '21: ACM Symposium on ACM Symposiums}{June 03--05, 2021}{Online, Earth}
\acmBooktitle{ACM '21: ACM Symposium on ACM Symposiums, June 03--05, 2021, Online, Earth}
\acmPrice{15.00}
\acmISBN{978-1-4503-XXXX-X/18/06}
\fi

\title{Some Lessons Learned Running Virtual Reality Experiments Out of the Laboratory}

\ifdefined\ARXIV

\author{Anthony Steed, Daniel Archer, Ben Congdon, Sebastian Friston, David Swapp, Felix J. Thiel\\
Department of Computer Science\\University College London\\
\texttt{Contact: A.Steed@ucl.ac.uk}
}
\else
\author{Anthony Steed\footnote{Contact: A.Steed@ucl.ac.uk}, Daniel Archer, Ben Congdon, Sebastian Friston, David Swapp, Felix J. Thiel}

\affiliation{%
  \institution{Department of Computer Science\\University College London}
  \streetaddress{66-72 Gower St}
  \city{London}
  \country{UK}
  \postcode{WC1E 6EA}
}

\fi

\ifdefined\ARXIV
\else

\fi

\begin{document}

\ifdefined\ARXIV
\maketitle
\else

\fi

\begin{abstract}
In the past twelve months, our team has had to move rapidly from conducting most of our user experiments in a laboratory setting, to running experiments in the wild away from the laboratory and without direct synchronous oversight from an experimenter. This has challenged us to think about what types of experiment we can run, and to improve our tools and methods to allow us to reliably capture the necessary data. It has also offered us an opportunity to engage with a more diverse population than we would normally engage with in the laboratory. In this position paper we elaborate on the challenges and opportunities, and give some lessons learned from our own experience.
\end{abstract}





\ifdefined\ARXIV
\else
\maketitle

\fi

\section{Introduction}

Our laboratory at University College London has a long history of running user experiments in virtual reality (VR), augmented reality and other novel user interfaces. Over the years, we have built up significant infrastructure for running user studies in the lab, but at the current time, we cannot access our labs. 


Fortunately, we are at a cusp in the commercialisation of VR and associated technologies. There are now millions of consumer VR systems out there, and this gives us an opportunity to run studies out of the lab. In this position paper we elaborate on our evolving practice in developing experiments out of the laboratory, but also discuss our rationale for proposing to continue at least part of our work in this manner going forward.



\section{Challenges and Opportunities}

There are a three ways that we have run or are planning to run studies out of the laboratory. The first is just publishing an application online. In this case it must run on consumer equipment (e.g. \cite{steed_wild_2016}). The second is a small variant, in that we publish an application but have other experimenters run it because the equipment is largely constrained to research labs (e.g. on relatively rare systems such on HoloLens, see \cite{steed_evaluating_2020}). The third is sending out specific sets of equipment to users, including, for example, a consumer HMD with some other tracking equipment (e.g. \cite{moustafa_longitudinal_2018}).

\subsection{Experiment Scope}

The main challenge in running studies remotely is the difficulty of supporting specific custom interface or capture technologies. For example, some of our experiments require one-off systems (e.g. \cite{steed_docking_2020} which uses a custom, and delicate, haptic system) or use combinations of monitoring equipment (e.g. \cite{yuan_is_2010} is one of several studies that use GSR). 
While we can lend out additional monitoring devices on their own or with consumer VR equipment, the main implication of this restriction is that experiments are mostly constrained to work within the limitations of the existing interface hardware. 

\subsection{Experiment Protocols}

The second challenge is that the protocols have to be self-running, self-documenting and robust to user behaviour. 
We have had to think about internal validity and how extra care is necessary to ensure that "unsupervised" experiment trials proceed in as uniform a way as possible.  In the lab, the entire process, from the moment a participant arrives until they leave at the end of the session, is normally highly regimented e.g. we offer them water, discuss the experiment, answer any questions they have about participant information and get their consent. With the "out of the lab" model, participants can still ask these questions (via email) but are less likely to. Experiment designs need to be tailored to reduce the scope for errors of misunderstanding since we cannot directly observe the participant as they go through any pre-trial tutorial phase.  In the lab each trial takes place in the same (uncluttered, reasonably spacious and quiet) space, whereas we have no control over the space the remote participant has.


\subsection{Ecological Validity}

The main opportunity for distributed experiments is that we can reach a different participant pool which is potentially more diverse. In the lab, although we use various participant pools and recruitment services, most participants are students or friends of students. They have specific motivations: they want to try VR, their friend tried it, they have a spare hour between classes or they get paid.

Out of the lab there are different motivations and we should acknowledge some potential biases. Participants own their hardware so will have some experience with virtual reality and games/social experiences. Participants could be biased towards being early adopt-ers, the population might skew in age or gender.
However, for certain types of experiment we desire participants who are already VR (or AR) users; we no longer plan any experiment where we expect participants to be naive. 

\subsection{Reproduction and Openness}

A second opportunity, and one that motivated some of our early efforts in this area, is that sharing experiments can enable people to understand our experiments and reproduce them, because they can experience them. Our Bar Experiment was available on a number of platforms for a few years, but those platforms (Gear VR, Cardboard) are now deprecated \cite{steed_wild_2016}. A slightly different angle on this was not a distributed experiment, but used commercial content, in that the experiment was based on an experience produced by the BBC which is still available, "We Wait". Aside from making experiments available, we hope to start sharing full experiment code. This is partly enabled by Microsoft open sourcing the RocketBox avatars \cite{gonzalez-franco_rocketbox_2020} as many of our demonstrations use those assets.



There is also a larger opportunity: recruitment of very large numbers of participants allowing different types of experiment with more conditions, or more open-ended participant engagement. To reach this, we obviously need to make the experiment attractive to run on its own; it is unlikely that this would be compensated.



\section{Technical Lessons}

In this section we cover, in no particular order, some insights from our experience of running experiments out of the lab. 

We have adopted two strategies to ensure participants are properly instructed and give informed consent: fully on the web as part of the download experience or from within the app. 
Our ethics approval allows for short-form instruction and consent in-app as long as participants can also access a long-form version online if they wish.

Our apps are designed to be easy to use. They must include their own tutorials .
We use standard interaction techniques such as teleportation and grabbing, and do not overload the user with instructions. User interfaces, especially if they involve questionnaires or other data entry need extensive testing. We direct participants to debriefing and often provide a simple summary of what happened within the applications at the end.

Distribution of applications to participants is not straightforward. On PC participants rely on our word that the application is not dangerous. We have made extensive use of the SideQuest platform (\cite{noauthor_sidequest_nodate}) to deliver applications because although it is not simple for end-users, the instructions are clear and they might appreciate having access to other demonstrations on that platform. 

A problem we have dealt with is compliance with the protocol. There are two main ways of dealing with this. In our Bar Experiment \cite{steed_wild_2016}, we put in various measures to monitor participants to make sure that they were active and looking in plausible directions. We also filtered the results to questionnaires to remove participants who answered too quickly or answered the same value for each question. We also balanced some questions, so the answers should have been opposite.

Many of the experiments have been designed to be used sat down. The reason for this is that we don't know the amount of space that is available to the user. While we could monitor the chaperone system, there is no guarantee that the participant has this correctly configured. 

We extensively test our applications and generally avoid anything that could cause simulator sickness. We have thus  avoided travel techniques whenever possible.

To satisfy data protection requirements we tend to log data to a secure server at UCL during the experience. No data is left on the device. However, we have adopted a process of keeping a simple count of the number of times the experiment has been run. We can't prevent a participant running the experiment multiple times, but we can ensure that we iterate through different conditions. Another workaround, especially when the use of additional peripheral devices is involved (such as for gathering biometric data), is to send a smartphone that pairs to the said device and stores the data via a custom mobile application.

A technical issue that we have fallen foul of and now know to avoid: make sure you use the correct clock to measure passage of time. Participants might stop the experiment midway (e.g. to answer the door), something they wouldn't do in the lab. At least Unity pauses the default timer when the HMD is removed, so logged time might not match elapsed absolute time. 
We advise participants that they should attempt to complete the experiment in one sitting, and we don't make the experiments too long. 

If participants are going to be paid, there needs to be a negotiation between a server and the application software. The easiest way we have found is that the application generates a unique code on completion that can be redeemed if emailed to us. This unique code can't be matched to the data we received.

Because we require participants to be online to log data, for some experiments, we have adopted a strategy of having a server distribute any necessary condition configuration so as to balance the number of participants in conditions.




\ifdefined\ARXIV
\section*{Acknowledgements}
\else
\begin{acks}
\fi

This work was partly funded by UK EPSRC project CASMS (grant reference EP/P004016/1) and EU H2020 project PRIME-VR2 (grant number 856998).

\ifdefined\ARXIV
\else
\end{acks}
\fi


\ifdefined\ARXIV
\bibliographystyle{unsrt}
\else
\bibliographystyle{ACM-Reference-Format}
\fi
\bibliography{sample-base}



\end{document}